# Ballistic conductivity of graphene channel with p-n junction on ferroelectric domain wall

*Anna N. Morozovska*[1,*], *Eugene A. Eliseev*[2], *and Maksym V. Strikha*[3,4,†],

[1] *Institute of Physics, National Academy of Sciences of Ukraine,*
*pr. Nauky 46, 03028 Kyiv, Ukraine*

[2] *Institute for Problems of Materials Science, National Academy of Sciences of Ukraine,*
*Krjijanovskogo 3, 03142 Kyiv, Ukraine*

[3] *Taras Shevchenko Kyiv National University, Radiophysical Faculty*
*pr. Akademika Hlushkova 4g, 03022 Kyiv, Ukraine*

[4] *V.Lashkariov Institute of Semiconductor Physics, National Academy of Sciences of Ukraine,*
*pr. Nauky 41, 03028 Kyiv, Ukraine*

**Abstract**

We study the impact of the ferroelectric domain wall on the ballistic conductance of the single-layer graphene channel in the heterostructure graphene / physical gap / ferroelectric film using Wentzel-Kramers-Brillouin approximation. Both self-consistent numerical modeling of the electric field and space charge dynamics in the heterostructure and approximate analytical theory show that the domain wall contact with the surface creates p-n junction in graphene channel. We calculated that the carriers' concentration induced in graphene by uncompensated ferroelectric dipoles originated from the spontaneous polarization abrupt near the surface can reach the values of $10^{19}$ m$^{-2}$ order, which is in two orders higher than it can be obtained for the gate doped graphene on non-ferroelectric substrates. Therefore we predict that graphene channel with the p-n junction caused by ferroelectric domain wall would be characterized by rather high ballistic conductivity.

---

[*] corresponding author, e-mail: anna.n.morozovska@gmail.com

[†] corresponding author, e-mail: maksym_strikha@hotmail.com

## 1. Introduction

Starting from the discovery by Geim and Novoselov [1, 2, 3] up to date graphene does not stop amazing researchers by its wonderful physical properties per se and unique functionalities of graphene-based heterostructures [4], where graphene channel is combined with different substrates in [5, 6]. For instance ferroelectric substrates with high permittivity can induce in graphene higher carrier concentration than traditional high-k substrates for the same gate voltages [7, 8].

Pronounced memory effects are induced by ferroelectric polarization reversal in gate-controlled nonvolatile graphene-on-ferroelectric memory [9, 10], as well as unusual resistance hysteresis in the field effect transistors based on multi-layer graphene-on-ferroelectric (GFeETs) [11]. This enables the construction of graphene-on-ferroelectric hybrid devices for low-voltage [12] and flexible transparent [13] electronics elements, bi-stable memory based on single-layer graphene-on-ferroelectric [14], organic ferroelectric-on-graphene based memories [15]. The scientific interest to graphene-on-ferroelectric increases permanently, the current state-of-art is reflected in several high-profile papers devoted to the studies of dynamic hysteresis of electric dipoles in GFeETs [16], ferroelectricity driven spatial modulation of carrier density in graphene [17], graphene-ferroelectric meta-devices for nonvolatile memory and logic-gate operations [18], and several topical reviews [19, 20, 21].

Charge carriers accumulation and/or trapping inevitably takes place at the graphene/ferroelectric interface [22]. The charge carriers screen the depolarization electric field induced in graphene by discontinuity of spontaneous polarization at the ferroelectric surface and can cause of versatile phenomena (as predicted theoretically), such as carrier density modulation induced by pyroelectric effect at graphene/ferroelectric interface [23], triggering of the space charge modulation in multi-layer graphene by ferroelectric domains [24] and unusual conductivity effects in graphene channel on organic ferroelectric substrate [25]. Note, that finite size effects can influence strongly on the nonlinear hysteretic dynamics of storied charge and electro-resistance in the multi-layer graphene-on-ferroelectric, at that the domain stripes of different polarity can induce domains with *p* and *n* conductivity and for the case the dominant carrier scattering mechanism can be randomly distributed *p-n*-junction potentials [26].

At the moment there are a lot of studies of p-n junctions in graphene. The junction was firstly experimentally realized by Williams et al [27] in a very simple system consisting of conventional gated graphene channel and a top gate that is superimposed through the dielectric layer above the right side of the channel. Under the changing the ratio of the voltages on these two gates one can create the doping by the electrons and holes of the left and right sides of the channel, at that creating p-n junction. Hopefully, the idea can be used for creating of graphene bipolar transistor technology (in fact not implemented yet). The conventional theory of the p-n junction resistance was proposed in the paper [28].

The most intriguing result obtained in [28] is a large electric field in the p-n junction in comparison with elementary estimations due to the small screening ability of two-dimensional (2D) Dirac quasiparticles, therefore, the resistance of this transition is quite small. Physical principles of such junctions operation are explained in review [29]. Many experimental studies have been devoted to the quantum Hall effect, the Klein (or Landau-Zener) tunneling [30] and Veselago lensing [31] in graphene p-n junctions.

However, the usage of multiple gates or chemical doping of separate parts of graphene channel surface remained sole experimental methods for design of p-n junction for a long time. Hinnefeld et al [32] created the p-n junction in graphene using the ferroelectric Pb(Zr,Ti)O3 (PZT) substrate, some part of which was screened by conventional SiO2. They obtained an interesting experimental dependence of conductivity on the gate voltage that has two minima corresponding to the Fermi level coincidence with electro-neutrality point in the electrons and holes area under the presence of p-n junction.

Our principal idea is that if graphene could be imposed on the ferroelectric domain wall, a p-n-transition can occur without any additional gates, doping, screening, etc. Below we consider theoretically the possibility, model the electric field and ballistic transport in a single-layer graphene channel placed on 180-degree ferroelectric domain wall. The characteristic feature of the graphene channel on ferroelectric domain wall I-V characteristic should be two minima in the conductance dependence on the gate voltage, similarly to [32].

**2. Model of 2D-graphene channel on 180-degree ferroelectric domain wall**

The geometry of the considered problem is shown in the **Figure 1.** 2D-graphene layer (being a channel) is separated from a ferroelectric layer with 180-degree domain wall structure by ultra-thin dielectric layer (ultra-thin physical gap or ferroelectric dead layer). The ferroelectric layer is in ideal electric contact with the gate electrode. Below we describe the how we model each of the layers.

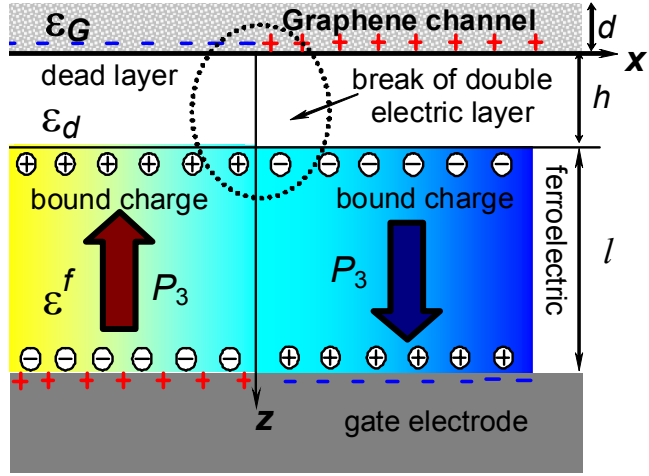

**Figure 1.** 180°-domain wall structure near the ferroelectric surface in the heterostructure graphene / dead layer / ferroelectric. The double electric layer is formed due to either the physical dead layer or intrinsic surface effect leading to diminished polarization at interface. Discontinuity of the double electric layer consisting of screening and bound changes results creates the depolarization electric field that stray into the gap.

**Single-layer graphene (2D-graphene channel).** We treat a single-layer graphene as a sheet of thickness $d$ of 3-5 Å order, for which we assume the validity of Thomas-Fermi approximation and introduce a screening length $R_S$. The assumption, although simple enough, can be approved by the first principle calculations of electronic density (see [17, 28]) and AFM measurements (see [33]). The problem of dielectric permittivity of 2D-graphene layer is still under debate (see e.g. [34]). It is strongly anisotropic, but, despite the previous estimations [24], where the value $\varepsilon_G^\perp \approx 15$ was usually taken in the graphene XY-plane, now it is generally accepted that $\varepsilon_G^\perp \approx 3$ and $\varepsilon_G^z \approx 1.5$ in normal z-direction.

**Dielectric layer.** Equation of state $\mathbf{D} = \varepsilon_0 \varepsilon_d \mathbf{E}$ relates the electrical displacement $\mathbf{D}$ and electric field $\mathbf{E}$ in the dielectric layer of thickness $h$ (either dead layer of ferroelectric or physical dielectric gap), $\varepsilon_0$ is a universal dielectric constant. The relative permittivity of the dielectric layer is $\varepsilon_d$, that is either equal to background permittivity of ferroelectric for a dead layer (5 – 10), or is about 1 for a physical gap. The potential $\varphi_d$ satisfies Laplace's equation inside the dielectric layer.

**Ferroelectric layer.** We consider a uniaxial ferroelectric of thickness $l$ with ferroelectric polarization $P_3^f$ directed along its polar axis z, with 180-degree domain wall – surface junctions [see **Fig. 1**]. Also we assume that the dependence of polarization components on the inner field

**E** can be linearized for transverse components as $P_1 = \varepsilon_0(\varepsilon_{11}^f - 1)E_1$ and $P_2 = \varepsilon_0(\varepsilon_{22}^f - 1)E_2$. Ferroelectric is dielectrically isotropic in transverse directions, i.e. relative dielectric permittivities are equal, $\varepsilon_{11}^f = \varepsilon_{22}^f$. Polarization z-component is $P_3(\mathbf{r}, E_3) = P_3^f(\mathbf{r}, E_3) + \varepsilon_0(\varepsilon_{33}^b - 1)E_3$, where a relative background permittivity $\varepsilon_{ij}^b \leq 10$ is introduced [35]. The ferroelectric permittivity $\varepsilon_{33}^f \gg \varepsilon_{33}^b$. Inhomogeneous spatial distribution of the ferroelectric polarization $P_3^f(z)$ will be determined from the Landau-Ginzburg-Devonshire (LGD) type Euler-Lagrange equations, $a_3 P_3 + a_{33} P_3^3 - g \dfrac{\partial^2 P_3}{\partial z^2} = E_3$ with the boundary conditions are of the third kind [36], $\left( P_3 \pm \lambda \dfrac{\partial P_3}{\partial z} \right)\bigg|_{z=h;L} = 0$. The physical range of extrapolation length $\lambda$ is 0.5 – 2 nm [37]. Constants $a_i$ and $a_{ij}$ are the coefficients of LGD potential expansion on the polarization powers (also called as linear and nonlinear dielectric stiffness coefficients). Quasi-static electric field is defined via electric potential as $E_3 = -\partial \varphi_f / \partial x_3$. The potential $\varphi_f$ satisfies Poisson equation inside a ferroelectric layer.

Hence, for the problem geometry shown in the **Figure 1** the system of electrostatic equations acquires the form:

$$\Delta \varphi_a = 0, \quad \text{for } -\infty < z < -d, \quad \text{(air or vacuum)} \quad (1a)$$

$$\left( \varepsilon_G^z \frac{\partial^2}{\partial z^2} + \varepsilon_G^\perp \Delta_\perp \right) \varphi_G - \frac{\varphi_G}{R_S^2} = 0, \quad \text{for } -d < z < 0, \quad \text{(graphene)} \quad (1b)$$

$$\Delta \varphi_d = 0, \quad \text{for } 0 < z < h, \quad \text{(dead layer or dielectric gap)} \quad (1c)$$

$$\left( \varepsilon_{33}^b \frac{\partial^2}{\partial z^2} + \varepsilon_{11}^f \Delta_\perp \right) \varphi_f = \frac{1}{\varepsilon_0} \frac{\partial P_3^f}{\partial z}, \quad \text{for } h < z < L. \quad \text{(ferroelectric)} \quad (1d)$$

3D-Laplace operator is $\Delta$, 2D-Laplace operator is $\Delta_\perp$. Boundary conditions to the system (1) are conventional; the continuity of the electric potential and normal component of displacement $D_3 = \varepsilon_0 E_3 + P_3$ at each interface (air / graphene, graphene / dielectric and dielectric / ferroelectric) and potential vanishing in air, $\varphi_a(x, y, z \to -\infty) = 0$ and at the bottom gate, $\varphi_f(x, y, L) = 0$.

In Equation (1b) the screening length $R_S$ has the sense of a "bare" Thomas-Fermi screening radius of 2D graphene [5, 38], that can be estimated as $R_S = \dfrac{\pi \varepsilon_0 \hbar v_F}{e^2 k_F}$, where $v_F$ is Fermi velocity of electrons in 2D-graphene, $\hbar$ is the Plank constant and $k_F = \sqrt{\pi n_{2D}}$ is the Fermi momentum. Note that the "dressed" screening radii become strongly anisotropic in 2D graphene, namely $R_S^\perp = \dfrac{\pi \varepsilon_0 \varepsilon_G^\perp \hbar v_F}{e^2 k_F}$ and $R_S^z = \dfrac{\pi \varepsilon_0 \varepsilon_G^z \hbar v_F}{e^2 k_F}$.

### 3. Results of the self-consistent numerical calculations

Results of the self-consistent numerical calculations of polarization distribution, electric potential and space charge are shown in the **Figure 2**. Designations and numerical values of parameters and constants used in our calculations are listed in the **Table SI** in the Suppl. Mat. [39].

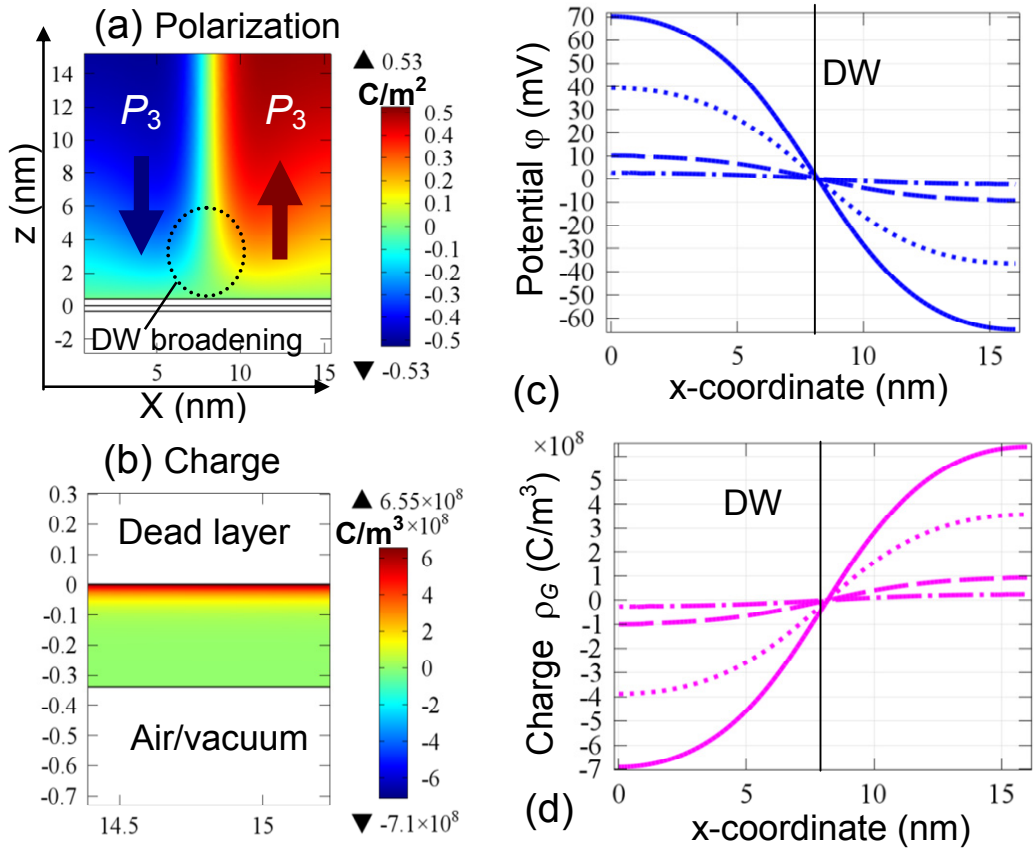

**Figure 2.** Electric potential and space charge distributions caused by the domain wall in the heterostructure graphene / dead layer / ferroelectric PZT. The distributions are calculated self-consistently. **(a)** Polarization distribution in ferroelectric calculated near its surface $z=h$. **(b)** The space charge distribution in a graphene layer calculated far from the domain wall – surface junction at x ≈ 15 nm. X-profiles of electrostatic potential **(c)** and space charge **(d)** at graphene surface $z=0$ (solid curves) and in its depth $z= -0.2$ Å, $-0.4$ Å and $-0.6$ Å (dotted, dashed and dashed-dotted curves).

**Figure 2a** shows the polarization distribution calculated self-consistently in ferroelectric near its surface $z = h$. A pronounced broadening of the domain wall is clearly seen. **Figure 2b** shows corresponding space charge distribution in a graphene layer far from the domain wall – surface junction. **Figures 2c** and **2d** show the rapid vanishing of the electrostatic potential φ and space charge ρ values from graphene surface $z=0$ (solid curves) towards its depth $z < 0$ (dotted, dashed and dashed-dotted curves). As anticipated φ and ρ profiles are anti-symmetric with respect to the wall plane $x=0$; they change sign at the wall plane $x = 0$ and saturate far from the wall. Normal and lateral components of electric field calculated at the graphene surface $z = 0$ and in its depth are shown in **Appendix A** of [39]. As anticipated $E_z$ x-profile is anti-symmetric with respect to the wall plane $x=0$, while $E_x$ profile is symmetric with respect to the wall plane $x=0$ and reaches a maximum at the plane.

### 4. Analytical model for the electric field and space charge distribution

In order to obtain analytical expression for graphene conductance and resistance one requires deriving analytical expressions for the space charge distribution and dynamics. Based on the numerical results shown in **Figures 2b-d**, we can assume that all uncompensated polarization bond charges are localized in thin sub-surface layer and considering the model case of abrupt domain wall, $P_3(x) = P_S \, \text{sign}(x)$, valid for a relatively thick ferroelectric layer. Thus, following ref.[40], the electric potential can be approximated by analytical expression in the dead layer/graphene interface region $0 \le z \le h$:

$$\varphi(x,z) = \frac{(P_S/\varepsilon_0)}{\varepsilon_{33}^f + \gamma \varepsilon_d} \frac{2\gamma}{\pi} \left( \begin{array}{l} \frac{x}{2} \ln\left( \frac{x^2 + (z-h^*)^2}{x^2 + (3h^* - z)^2} \right) + (z - h^*) \cdot \arctan\left( \frac{x}{z - h^*} \right) \\ -(3h^* - z)\arctan\left( \frac{x}{3h^* - z} \right) \end{array} \right) \quad (2)$$

The anisotropy factor $\gamma = \sqrt{\varepsilon_{33}^f / \varepsilon_{11}^f}$ is within the range 0.1 – 1 for typical ferroelectric materials. The spontaneous polarization $P_S$ is typically about (0.1 – 1) C/m². Parameter $h^*$ is the effective screening charge-surface separation, which is equal to $h^* = h + (\varepsilon_d/\varepsilon_G^z)R_S^z \equiv h + \varepsilon_d R_S$ [41]. Expression (2) is a good approximation if the ferroelectric layer thickness $l$ is much higher than $h^*$. Expressions for electric field components are listed in **Appendix A** of Suppl. Mat [39].

Expression (2) allows one to estimate the 2D-carrier concentration $n_{2D}$ in a graphene sheet far from the domain wall, as the space charge concentration integrated over graphene thickness $d$, $n_{2D}(x) = \frac{1}{e}\int_{-d}^{0} \rho_G(x,z)dz \sim \int_{-d}^{0} \frac{\varepsilon_0 \varepsilon_{33}^f}{eR_S^2}\varphi_G(x,z)dz$. In accordance with Gauss theorem the integration gives the electric displacement at the surfaces of graphene layer, $D_z(x,0) - D_z(x,-d)$. Since for the considered case a complete screening of electric potential in graphene takes place in the immediate vicinity of its surface (see dashed-dotted curve in the **Figures 2c** and **2d**), $D_z(x,-d) \to 0$ and $D_z(x,0) = \varepsilon_0 \varepsilon_{33}^f E_z(x,0)$, the upper estimation for the carrier concentration far from the domain wall ($x \to \infty$),

$$n_{2D}(x) = \frac{\varepsilon_0 \varepsilon_{33}^f}{e} |E_z(x,0)| \underset{x \to \pm\infty}{\cong} \frac{2\gamma \varepsilon_{33}^f (P_S/e)}{\varepsilon_{33}^f + \gamma \varepsilon_d}. \qquad (3)$$

Equation (3) means that the free carriers' concentration in graphene is proportional to the electric field distribution $E_z(x,0)$ as anticipated. Note that the concentration (3) becomes thickness dependent under the ferroelectric layer thickness decrease, because the spontaneous polarization $P_S$ and relative permittivity $\varepsilon_{33}^f$ become thickness dependent. Actually, when ferroelectric layer thickness $l$ approaches the critical thickness $l_{cr} = (10 - 100)$ nm the ferroelectric polarization decreases according to expression $P_S = P_S^{bulk} \sqrt{1 - l_{cr}/l}$ and then disappears at $l < l_{cr}$ [42]. Static dielectric permittivity "longitudinal" component $\varepsilon_{33}^f$ is thickness dependent, $\varepsilon_{33}^f(l) = \varepsilon_{33}^{bulk}/(1 - l_{cr}/l)$. The thickness dependence of the "transverse" component $\varepsilon_{11}^f$ is almost absent for uniaxial ferroelectrics; it is described by expression $\varepsilon_{11}^f(l) = \varepsilon_{11}^{bulk}/(1 - l_{cr}^\perp/l)$ for multiaxial ones. Due to the absence of depolarization effect in the ferroelectric layer plane the thickness $l_{cr}^\perp$ is typically much less than $l_{cr}$. Thus the thickness dependence of the anisotropy factor γ can be approximated as $\gamma \approx \gamma^{bulk}/\sqrt{1 - l_{cr}/l}$, where $\gamma^{bulk} \approx \sqrt{\varepsilon_{33}^{bulk}/\varepsilon_{11}^{bulk}}$.

**Figure 3** shows the electric potential φ and field $E_{x,z}$ distributions calculated analytically at the domain wall in the heterostructure graphene/dead layer/ferroelectric PZT. φ and $E_z$ are zero, while $E_x$ has a maximum at the wall plane $x = 0$. The small differences between the profiles of φ, $E_x$ and $E_z$ calculated analytically at z=0 for abrupt polarization distribution [shown in the **Figures 3** by solid curves] and the profiles, calculated numerically in a self-consistent way [shown in the **Figures 2c** and **2d** by solid curves], originate from the broadening of the domain wall at the wall-surface junction [shown in the **Figure 2a**]. Despite the wall broadening are noticeable, the differences between analytical and numerical results are very small inside the graphene single-layer at $z \leq 0$, that justifies the validity of approximate analytical expressions (2)-(3).

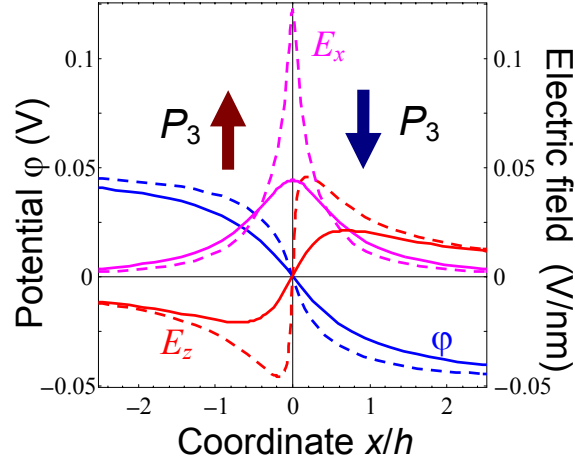

**Figure 3.** Electric potential and field distribution along the hetero-interface caused by the domain wall (located at x=0) in the heterostructure graphene/dead layer/ferroelectric PZT. Distributions are calculated analytically in the abrupt domain wall approximation, $P_3(x) = P_S \, \mathrm{sign}(x)$. Coordinate $z_2= 0$ corresponds to the dead layer/graphene interface (solid curves) and $z_1= +0.99h$ corresponds to the dead layer/ferroelectric (dashed curves).

## 5. Domain wall surface-junction creates a p-n junction in graphene channel

Let us calculate the probability $w$ for electron, starting from *n*-region with a wave vector $k$ directed at $\vartheta$ angle from *x* axis, to pass into *p*-region (similarly to the way it was done in [43]). In the center of the junction, at x=0, electron's kinetic energy is $v_F \hbar \sqrt{k_x^2 + k_y^2}$, where the *y*-component of momentum, $k_y = k_F \sin\vartheta$, is conserved. Therefore *x* component is determined by expression $k_x(x) = \sqrt{(e\varphi(x,0)/v_F\hbar)^2 - (k_F \sin\vartheta)^2}$. The classically allowed region of electron's motion is determined by inequality $e\varphi(x,0) > v_F \hbar k_y$, which means that the electron cannot overpass the turning point at distance $l_x$ from the center of the p-n-junction located at the domain wall plane $x = 0$.

Under the condition $\sin\vartheta = 0$, corresponding to the movement strictly along *x* axis, as well for the small angles $\vartheta \ll 1$, the probability $w$ can be estimated quasi-classically in the Wentzel–Kramers–Brillouin (WKB) approximation:

$$w \approx e^{-2S/\hbar}, \qquad S = i\hbar \int_{-l_x}^{l_x} k_x(x)dx \qquad (4)$$

In order to perform integration in Eq.(4), one can use the Pade approximation of electric potential (1b), $e\varphi(x,0) \approx v_F k_F \hbar Q \theta(x/h^*)$, where the odd function $\theta(y)$ has a simple piecewise smooth approximation $\theta(|y| < \pi/\ln 3) = y$ and $\theta(|y| > \pi/\ln 3) = \pm\pi/\ln 3$ ($\ln 3 \approx 1.09$). The polarization dependent dimensionless factor $Q$ can be derived from Eq.(2) as

$$Q = \frac{e(P_S/\varepsilon_0)}{\varepsilon_{33}^f + \gamma\varepsilon_d} \frac{2\gamma h^* \ln 3}{\pi\hbar k_F v_F}. \qquad (5)$$

Using Eq.(3) and the definition of parameter $Q$, Eq.(5), the expression for the Fermi wave vector has the form:

$$k_F = \sqrt{\pi n_{2D}} \equiv \frac{\pi^2 \varepsilon_0 \varepsilon_{33}^f}{e^2} \frac{\hbar v_F Q}{h^* \ln 3}. \qquad (6)$$

The distance $l_x$ is defined from the condition $e\varphi(l_x, 0) = v_F \hbar\, k_F \sin\vartheta$. Using anzatz $e\varphi(l_x, 0) \approx v_F k_F \hbar Q l_x / h^*$, one elementary derives that $l_x = h^* \sin\vartheta / Q$. For the angle $\vartheta = 0$ the classically forbidden region vanishes at all, and so $w(0) = 1$. Substitution of the linear approximation for $\theta(|y| < \pi/\ln 3) = y$ in the range of $\{-l_x, l_x\}$ for the small angles $\vartheta$, yields in Eq.(4): $S = (\hbar P_x/2)\sin^2\vartheta$, $w(\vartheta) \cong \exp(-P_x \sin^2\vartheta)$, where the probability factor is $P_x \approx \pi^3 \varepsilon_0 \varepsilon_{33}^f \hbar v_F / (e^2 \ln 3)$. Since the estimation for the exponential factor $P_x$ is typically essentially higher than unity (more than 100 for $\varepsilon \geq 500$), conduction per unit length of the p-n junction induced by a domain wall can be estimated as:

$$\sigma_{pn} = \frac{e^2}{\pi\hbar} \int_{-\vartheta_0}^{\vartheta_0} \frac{k_F w(\vartheta)}{2\pi} d\vartheta \approx \frac{e^2 k_F}{2\pi^2\hbar} \sqrt{\frac{\pi}{P_x}} Erf(\sqrt{P_x}\vartheta_0). \qquad (7)$$

Since the wave vector $k_F \equiv \sqrt{\pi n_{2D}}$ in accordance with Eq.(8), the probability factor $P_x \approx \pi^3 \varepsilon_0 \varepsilon_{33}^f \hbar v_F / e^2$ and the angle $\vartheta_0 \approx 1/\sqrt{P_x}$ is small enough, for estimates Eq.(7) can be simplified to the form:

$$\sigma_{pn} \cong \frac{e^2}{\pi\hbar} \sqrt{\frac{\alpha}{\varepsilon_{33}^f} \frac{c}{\pi v_F}} n_{2D} \qquad (8)$$

In Equation (8) we introduced the fine-structure constant $\alpha = e^2/4\pi\varepsilon_0(\hbar c)$, $c$ is the light velocity in vacuum. Since $P_x \sim v_F$ and $n_{2D} \sim |E_z|$, Eq.(8) means that $\sigma_{pn} \sim \sqrt{|E_z|/v_F}$.

Note, that Eq.(8) coincides (with accuracy of the dimensionless factor) to a well known expression for ballistic conductivity of graphene, $\sigma_{pn} = 2e^2\sqrt{n_{2D}}/\hbar$ [21, 44]. However, Eq.(8) includes a factor $\sqrt{\alpha c/(\varepsilon_{33}^f \pi v_F)} \sim 0.1$. On the other hand, the estimation for concentration caused by ferroelectric dipoles, leads to values of 1019 m-2 order, which is in two orders higher, than it can be obtained for the gate doped graphene on mica substrate (see e.g. [8]). Therefore, the graphene p-n junction on ferroelectric domain wall would be characterized by rather high ballistic conductivity and low resistance correspondingly. A nominal resistance $R_{np}$ can be written as, $R_{np} = 1/(a\sigma_{np})$, where $a$ is p-n junction width.

Expression (8) differs from Eq.(2) in [43], because in our case $n_{2D}$ is defined by a spontaneous polarization of ferroelectric in accordance with Eq.(3). Conductivity and resistance redistributions caused by the domain wall are shown in the **Figure 4a**. 3D plot of the conductance is shown in the **Figure 4b**. As one can see the width of the p-n junction region is about $2h$.

Since $\sigma_0 = e^2/(2\pi\hbar)$ is an elementary quantum conductance, the ratio $\sigma_{pn}/\sigma_0$ gives us the notion about the amount of conductance modes. This result corresponds in order of values with one, calculated in [45]. **Figure 4c** illustrates the dependences of the dimensionless conductance $\sigma_{pn}/\sigma_0$ and resistance $\sigma_0/\sigma_{pn}$ on the spontaneous polarization value *PS*. As one can see from the figure the amount of conductance modes (that is already about 107 at $Ps$ = 0.5 C/m2) increases with *PS* increase as $\sigma_{pn}/\sigma_0 \sim \sqrt{P_S}$.

Finally let's study finite size effect of graphene conductance and resistance appeared under the ferroelectric film thickness decrease. Since the expression (3) for free carriers concentration $n_{2D}$ become thickness dependent under the ferroelectric film thickness $l$ decrease, because $P_S = P_S^{bulk}\sqrt{1-l_{cr}/l}$, $\varepsilon_{33}^f(l) = \varepsilon_{33}^{bulk}/(1-l_{cr}/l)$ and $\gamma \approx \gamma^{bulk}/\sqrt{1-l_{cr}/l}$, the conductance thickness dependence acquires the form:

$$\sigma_{pn} \cong \frac{e^2}{\pi\hbar}\sqrt{\frac{\alpha c}{\pi v_F}\frac{2\gamma^{bulk}(P_S^{bulk}/e)(1-l_{cr}/l)}{\varepsilon_{33}^{bulk} + \gamma^{bulk}\varepsilon_d\sqrt{1-l_{cr}/l}}} \quad \text{at } l > l_{cr}. \qquad (9)$$

We concluded from the expression (9) that the conductance modulation by domain structure monotonically decreases under the thickness $l$ decrease according to the law $\sigma_{pn} \sim \sqrt{1-l_{cr}/l}$ and becomes zero at $l \leq l_{cr}$. At the same time resistance increases under the $l$ decrease as $R_{pn} \sim 1/\sqrt{1-l_{cr}/l}$. Dependences of the dimensionless conductance $\sigma_{pn}/\sigma_0$ and resistance $\sigma_0/\sigma_{pn}$ on the reduced thickness of ferroelectric film $l/l_{cr}$ is shown in the **Figure 4d**.

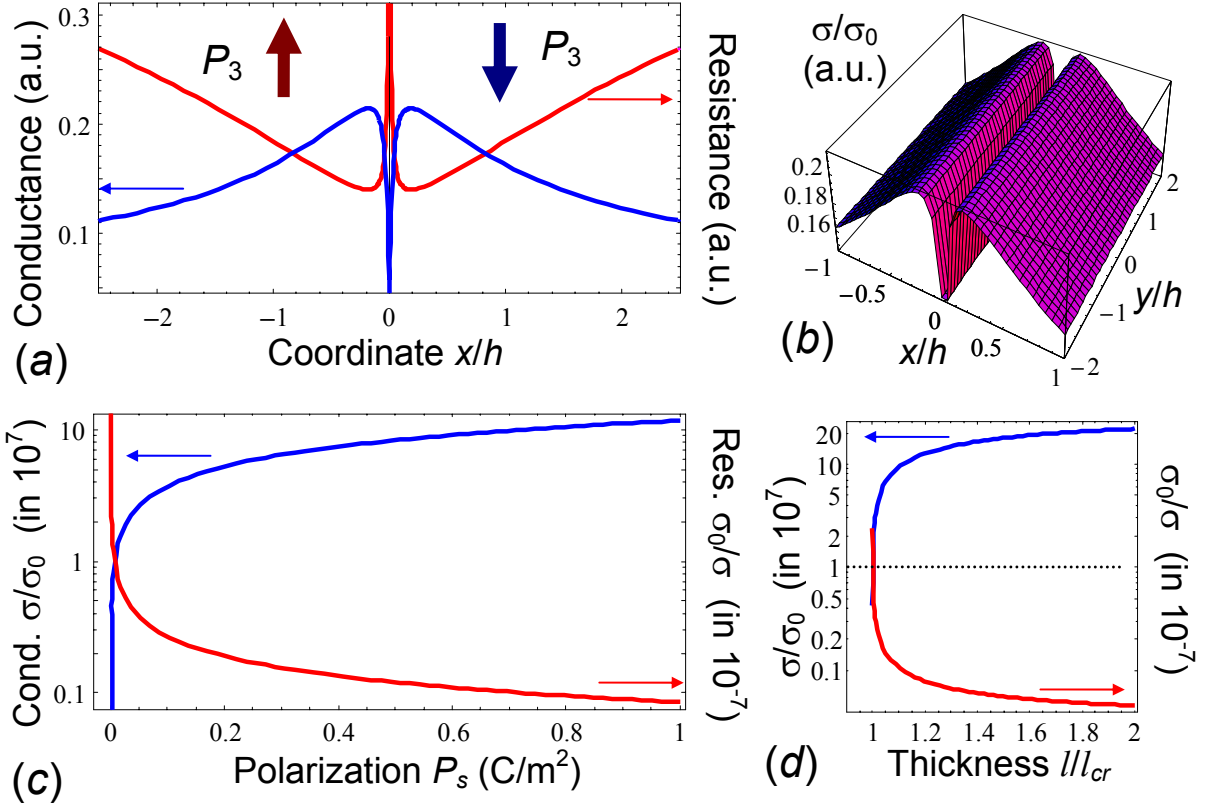

**Figure 4.** (a) Conductance (blue curves) and resistance (red curves) x-distributions along the heterointerface caused by the domain wall located at x=0 in the heterostructure graphene/dead layer/ferroelectric PZT. Parameters $P_S$=0.7 C/m$^2$, $\gamma$= 0.8, $\varepsilon_{11}^f=\varepsilon_{33}^f$=500, $\varepsilon_g$= 1, $h$= 0.4 nm. (b) 3D plot of the p=n junction conductance. (c, d) Dependences of the dimensionless conductance $\sigma_{pn}/\sigma_0$ (multiplied by factor 10$^{-7}$) and resistance $\sigma_0/\sigma_{pn}$ (multiplied by factor 10$^{+7}$) on the spontaneous polarization value (c) and reduced thickness of ferroelectric film $l/l_{cr}$ (d).

Note also, that the multi-particle treatment [28], which corrects essentially the results of ref.[43], cannot be applied for the considered case, because the charge carriers in ferroelectric are fixed and their distribution is determined by the problem geometry only.

## Conclusion

Using WKB approximation we have studied the impact of the ferroelectric domain walls on the ballistic conductance of the single-layer graphene channel in the heterostructure graphene / physical gap / ferroelectric film. Both self-consistent numerical modeling of the electric field and space charge dynamics in the system and approximate analytical theory show that the domain wall creates p-n junction in graphene channel. We had calculated the impact of the ferroelectric domain on the ballistic conductance of the single-layer graphene channel. The amount of conductance modes increases under the spontaneous polarization $P_S$ proportionally to $\sqrt{P_S}$. The conductance modulation by domain structure monotonically decreases under the film thickness $l$ decrease according to the law $\sigma_{pn} \sim \sqrt{1 - l_{cr}/l}$ and becomes zero at $l \leq l_{cr}$. At the same time graphene resistance increases under the $l$ decrease as $R_{pn} \sim 1/\sqrt{1 - l_{cr}/l}$.

The principal result is that the carriers' concentration induced in graphene by uncompensated ferroelectric dipoles originated from the spontaneous polarization abrupt near the surface can reach the values of $10^{19}$ m$^{-2}$ order, which is in two orders higher than it can be obtained for the gate doped graphene on non-ferroelectric substrates. Therefore we predict that the p-n junction in graphene-on-ferroelectric domain wall would be characterized by rather high ballistic conductivity. Moreover the sequence of quasi-parallel domain walls will induce unusual cross-talk effects of graphene conductance. Since the domain walls can be perfectly pinned by the lattice relief and point defects, our theoretical predictions can be readily verified experimentally.

**Acknowledgements.** E.A.E. and A.N.M. acknowledge the Center for Nanophase Materials Sciences, which is a DOE Office of Science User Facility, project CNMS2016-061.

**Authors contribution.** M.V.S. formulated the research idea, physical model and performed analytical calculations of graphene conductance. A.N.M. formulated electrostatic problem and derived its analytical solution as well as study size effects in the system. E.A.E. performed self-consistent numerical simulations of the electric field and space charge dynamics in the heterostructure. A.N.M. and M.V.S. wrote the manuscript.

# Supplementary Materials

### Appendix A. Single-domain wall case

Results of the self-consistent numerical calculations of electric field components are shown in the **Figure S1**. Designations and numerical values of parameters and constants used in our calculations are listed in the **Table SI.**

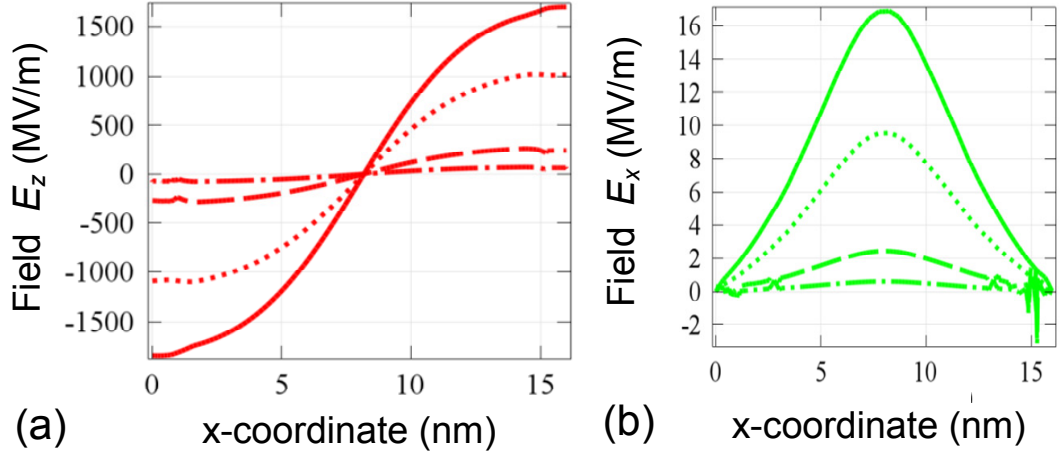

(a)    (b)

**Figure S1.** Electric field distributions caused by the domain wall (located at x=0) in the heterostructure graphene/dead layer/ferroelectric PZT. Plots are calculated self-consistently. X-profiles of the normal **(a)** and lateral **(b)** components of electric field calculated at graphene surface z=0 (solid curves) and in its depth z= −0.2 Å, −0.4 Å and −0.6 Å (dotted, dashed and dashed-dotted curves). Parameters are listed in the **Table I.**

**Table SI**. Parameters and constants designations and numerical values

| Parameter, constant or value | Numerical value and dimensionality |
|---|---|
| single-layer graphene thickness | $d \approx 3.4 \times 10^{-10}$ m (one lattice period of graphite) |
| dielectric (dead) layer thickness | $h = 4 \times 10^{-10}$ nm (one lattice period of ferroelectric) |
| universal dielectric constant | $\varepsilon_0 = 8.85 \times 10^{-12}$ F/m (e/Vm) |
| dielectric permittivity of graphene layer | $\varepsilon_G^\perp \approx 3$ and $\varepsilon_G^z \approx 1.5$ |
| permittivity of the dielectric layer | $\varepsilon_d = 1$ (typical range 1 – 10) |
| permittivity of the ferroelectric film | $\varepsilon_{33}^f = 500$, $\varepsilon_{11}^f = \varepsilon_{22}^f = 780$ |
| dielectric anisotropy of ferroelectric film | $\gamma = \sqrt{\varepsilon_{33}^f / \varepsilon_{11}^f} = 0.8$ |
| spontaneous polarization in a bulk | $P_S^{bulk} = (0.5 - 0.7)$ C/m² |
| Plank constant | $\hbar = 1.056 \times 10^{-34}$ J·s = $6.583 \times 10^{-16}$ eV·s |
| Fermi velocity of electrons in graphene | $v_F \approx 10^6$ m/s |
| Fermi momentum of electrons in graphene | $k_F = \sqrt{\pi n_{2D}} \sim \sqrt{\pi (P_S / e)} \approx 3.7 \times 10^9$ m⁻¹ |

| "bare" Thomas-Fermi (TF) screening radius | $R_S = \dfrac{\pi \varepsilon_0 \hbar v_F}{e^2 k_F} \approx 3 \times 10^{-11}$ m |
|---|---|
| "dressed" transverse TF screening radii | $R_S^\perp = \dfrac{\pi \varepsilon_0 \varepsilon_G^\perp \hbar v_F}{e^2 k_F} \approx 9 \times 10^{-11}$ m |
| "dressed" longitudinal TF screening radii | $R_S^z = \dfrac{\pi \varepsilon_0 \varepsilon_G^z \hbar v_F}{e^2 k_F} \approx 4.5 \times 10^{-11}$ m |
| carrier concentration | $n_{2D} \cong \dfrac{2\gamma \varepsilon_{33}^f (P_S/e)}{\varepsilon_{33}^f + \gamma \varepsilon_d} = 6{,}99 \times 10^{18}$ m$^{-2}$ |
| probability factor | $P_x = \dfrac{\pi^3 \varepsilon_0 \varepsilon_{33}^f \hbar v_F}{e^2 \ln 3} \approx \dfrac{\pi^3 \varepsilon_0 \varepsilon_{33}^f \hbar v_F}{e^2} > 100$ |
| elementary conductance | $e^2/(2\pi \hbar) = 3{,}9 \times 10^{-5}$ Siemens |
| fine-structure constant | $\alpha = e^2/4\pi \varepsilon_0 (\hbar c) \approx 1/137$ |

Assuming, that all uncompensated polarization bond charges are localized in thin near-surface layer and considering the model case of abrupt domain wall, $P_3(x) = P_S \, \text{sign}(x)$, in a relatively thick ferroelectric film, electric potential can be approximated by analytical expressions for the case $z \geq h$

$$\varphi(x,z) = \frac{(P_S/\varepsilon_0)}{\varepsilon_{33}^f + \gamma \varepsilon_d} \frac{2\gamma}{\pi} \left( \begin{array}{l} +\dfrac{x}{2}\ln\left(\dfrac{x^2 + ((z-h^*)/\gamma)^2}{x^2 + (2h^* + ((z-h^*)/\gamma))^2}\right) + \dfrac{z-h^*}{\gamma} \cdot \arctan\left(\dfrac{x}{(z-h^*)/\gamma}\right) \\ -\left(2h^* + \dfrac{z-h^*}{\gamma}\right)\arctan\left(\dfrac{x}{(z-h^*)/\gamma + 2h^*}\right) \end{array} \right) \quad \text{(A.1a)}$$

For the case $0 \leq z \leq h$:

$$\varphi(x,z) = \frac{(P_S/\varepsilon_0)}{\varepsilon_{33}^f + \gamma \varepsilon_d} \frac{2\gamma}{\pi} \left( \begin{array}{l} \dfrac{x}{2}\ln\left(\dfrac{x^2 + (z-h^*)^2}{x^2 + (3h^* - z)^2}\right) + (z - h^*) \cdot \arctan\left(\dfrac{x}{z-h^*}\right) \\ -(3h^* - z)\arctan\left(\dfrac{x}{3h^* - z}\right) \end{array} \right) \quad \text{(A.1b)}$$

Here $\varepsilon_{33}^f$ is the dielectric permittivity of the ferroelectric (including background or reference state $\varepsilon_{33}^b \leq 10$); $\varepsilon_0$ is the dielectric constant, $h^*$ is the effective screening charge-surface, which is expressed via the actual thickness of the dielectric gap $h$ and the screening length of graphene channel $R_d$ as $h^* = h + (\varepsilon_g/\varepsilon_S)R_d$. The anisotropy factor is introduced as $\gamma = \sqrt{\varepsilon_{33}^f/\varepsilon_{11}}$. The isotropic dielectric permittivity of dead layer is $\varepsilon_g$. Expressions (A.1) are good approximations if the ferroelectric film thickness $l$ is much higher than $h^*$.

The stray field vertical and radial components produced by the domain wall in the ferroelectric ($z > h$) are

$$E_z = -\frac{(P_S/\varepsilon_0)}{\varepsilon_{33}^f + \gamma\varepsilon_d} \frac{2}{\pi}\left(\arctan\left(\frac{x}{(z-h^*)/\gamma}\right) - \arctan\left(\frac{x}{2h^* + ((z-h^*)/\gamma)}\right)\right), \quad \text{(A.2a)}$$

$$E_x = -\frac{(P_S/\varepsilon_0)}{\varepsilon_{33}^f + \gamma\varepsilon_d} \frac{\gamma}{\pi}\ln\left(\frac{x^2 + (2h^* + ((z-h^*)/\gamma))^2}{x^2 + ((z-h^*)/\gamma)^2}\right). \quad \text{(A.2b)}$$

The stray field vertical and radial components produced by the domain wall in the dead layer ($0 \leq z \leq h$) are

$$E_z = -\frac{(P_S/\varepsilon_0)}{\varepsilon_{33}^f + \gamma\varepsilon_d} \frac{2\gamma}{\pi}\left(\arctan\left(\frac{x}{z-h^*}\right) + \arctan\left(\frac{x}{3h^* - z}\right)\right) \quad \text{(A.3a)}$$

$$E_x = -\frac{(P_S/\varepsilon_0)}{\varepsilon_{33}^f + \gamma\varepsilon_d} \frac{\gamma}{\pi}\ln\left(\frac{x^2 + (3h^* - z)^2}{x^2 + (z-h^*)^2}\right) \quad \text{(A.3b)}$$

Equations (A.3) are approximation of the series in the limiting case when the domain structure period tends to infinity.

**Figure S2** shows the electric potential and field distribution calculated analytically at the domain wall in the heterostructure graphene/dead layer/ferroelectric PZT. As anticipated φ and $E_z$ are zero, while $E_x$ has a maximum at the wall plane $x = 0$. The small differences between the profiles of φ, $E_x$ and $E_z$ calculated analytically at $z=0$ for abrupt polarization distribution [shown in the **Figures S2** by solid curves] and the profiles, calculated numerically in a self-consistent way [shown in the **Figures S2c** and **S2d** by solid curves], originate from the broadening of the domain wall at the wall-surface junction [shown in the **Figure S2a**]. Despite the wall broadening are noticeable, the differences between analytical and numerical results are very small inside the graphene single-layer at $z \leq 0$, that justifies the validity of approximate analytical expressions (A.2)-(A.3).

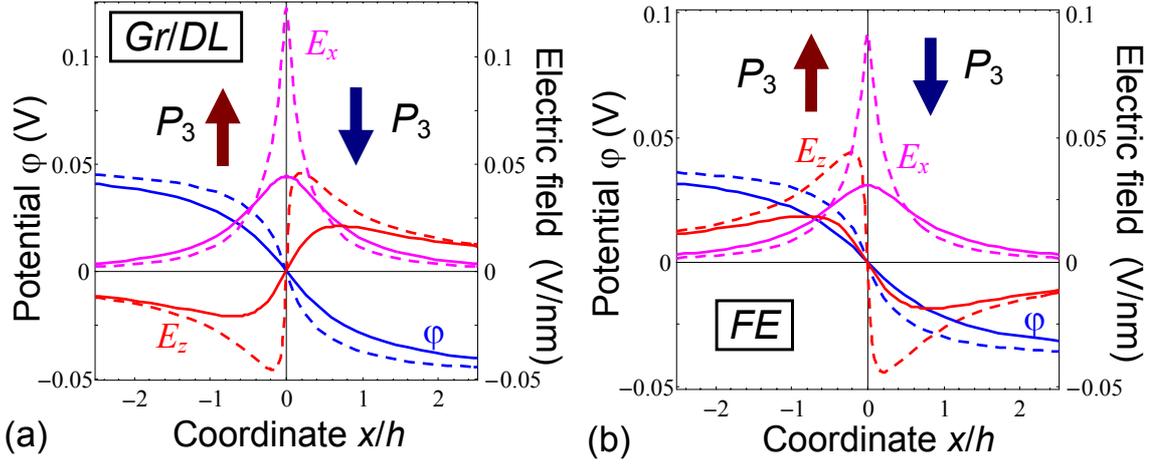

**Figure S2.** Electric potential and field distribution along the hetero-interface caused by the domain wall (located at x=0) in the heterostructure graphene/dead layer/ferroelectric PZT. Distributions are calculated analytically in the abrupt domain wall approximation, $P_3(x) = P_S \text{sign}(x)$. Plot **(a)**: coordinate $z_1 = +0.99h$ corresponding to the ferroelectric/dead layer interface (dashed curves) and $z_2 = 0$ corresponding to the dead layer/graphene interface (solid curves). Plot **(b)**: coordinate $z_1 = +1.01h$ corresponding to the ferroelectric surface (dashed curves) and $z_2 = +2h$ corresponding to the ferroelectric depth (solid curves). Parameters $P_S = 0.7$ C/m², $\gamma = 0.8$, $\varepsilon_{33}^f = 500$, $\varepsilon_d = 1$, $h = 0.4$ nm.

### Appendix B. Domain wall surface-junction creates a p-n junction in graphene

Let us calculate the probability $w$ for electron, starting from *n*-region with $k$ directed at $\vartheta$ angle from $x$ axis, to pass into *p*-region. In the center of the junction, at $x=0$, electron's kinetic energy is $v_F \hbar \sqrt{k_x^2 + k_y^2}$, where within the transition the $y$ momentum component $k_y = k_F \sin\vartheta$ is conserved. Therefore $x$ component is determined by:

$$k_x(x) = \sqrt{\left(\frac{e\varphi(x,0)}{v_F \hbar}\right)^2 - (k_F \sin\vartheta)^2}, \qquad (B.1)$$

and classically allowed region of electron's movement is determined by inequality $e\varphi(x,0) > v_F \hbar k_y$, which means that the electron can't overpass the turning point at distance $l_x$ from the center of the p-n-junction located at the domain wall plane $x = 0$.

At $\sin\vartheta = 0$, corresponding to the movement strictly along $x$ axis, as well for the small angles $\vartheta \ll 1$, the probability $w$ can be estimated quasi-classically in the WKB approximation:

$$w \approx e^{-2S/\hbar}, \qquad S = i\hbar \int_{-l_x}^{l_x} k_x(x)dx \qquad (B.2)$$

In order to perform integration in Eq.(B.2) one can use the Pade approximation of electric potential (1b)

$$e\varphi(x,0) \approx v_F k_F \hbar Q \theta(x/h^*) \qquad (B.3a)$$

Here polarization dependent dimensionless factor $Q$ is

$$Q = \frac{e(P_S/\varepsilon_0)}{\varepsilon_{33}^f + \gamma \varepsilon_d} \frac{2\gamma h^* \ln 3}{\pi \hbar k_F v_F} \approx \frac{e(P_S/\varepsilon_0)}{\varepsilon_{33}^f + \gamma \varepsilon_d} \frac{2\gamma h^*}{\pi \hbar k_F v_F}, \qquad (B.3b)$$

We used that $\ln 3 \approx 1.09$. The odd function $\theta(y)$ has a simple piecewise smooth approximation $\theta(|y| < \pi/\ln 3) = y$ and $\theta(|y| > \pi/\ln 3) = \pm \pi/\ln 3$. Its Pade approximation is $\theta(y) \approx \dfrac{-y}{1 + |y|\pi^{-1}\ln 3}$.

Fermi wave vector is determined in a conventional way as $k_F = E_F/\pi v_F \equiv \sqrt{\pi n_{2D}}$. Since $D_z = en_{2D}$ for the case a complete screening of spontaneous polarization, that gives the upper estimation for the carrier concentration far from the domain wall, $n_{2D} = \varepsilon_0 \varepsilon_{33}^f \dfrac{E_z(x \to \infty)}{e} \cong \dfrac{2\gamma \varepsilon_{33}^f (P_S/e)}{\varepsilon_{33}^f + \gamma \varepsilon_d}$. Using the definition of parameter $Q$, Eq.(B.3b) it yields that $\dfrac{2\gamma \varepsilon_{33}^f (P_S/e)}{\varepsilon_{33}^f + \gamma \varepsilon_d} = \varepsilon_0 \varepsilon_{33}^f \dfrac{\pi \hbar k_F v_F Q}{e^2 h^* \ln 3}$, and so $n_{2D} = \varepsilon_0 \varepsilon_{33}^f \dfrac{\pi \hbar k_F v_F Q}{e^2 h^* \ln 3}$, finally:

$$k_F = \frac{\pi^2 \varepsilon_0 \varepsilon_{33}^f}{e^2} \frac{\hbar v_F Q}{h^* \ln 3}, \qquad (B.4)$$

The distance is defined from the condition $e\varphi(l_x,0) = v_F \hbar k_F \sin \vartheta$. Using anzatz $e\varphi(l_x,0) \approx v_F k_F \hbar Q l_x/h^*$, we elementary derive that

$$l_x = \frac{h^*}{Q} \sin \vartheta \qquad (B.5)$$

For the angle $\vartheta = 0$ (movement strictly along $x$ axis) the classically forbidden region vanishes at all, and so $w(0) = 1$. However, for the small angles $\vartheta$, the linear approximation for $\theta(|y| < \pi/\ln 3) = y$ in the range of $\{-l_x, l_x\}$ yields in Eq.(B.2):

$$\frac{S}{\hbar} = k_F \int_{-l_x}^{l_x} \sqrt{\sin^2 \vartheta - \left(Q\frac{x}{h^*}\right)^2} \, dx =$$

$$= k_F l_x \left( \sqrt{\sin^2 \vartheta - \left(\frac{Q l_x}{h^*}\right)^2} + \sin^2 \vartheta \frac{h^*}{l_x Q} \arctan\left(\frac{(Q l_x/h^*)}{\sqrt{\sin^2 \vartheta - (Q l_x/h^*)^2}}\right) \right) \qquad (B.6)$$

Using Eqs.(B.4) and (B.5) in Eq(B.6), we obtain that $k_F l_x = \dfrac{\pi^2 \varepsilon_0 \varepsilon_{33}^f \hbar v_F}{e^2 \ln 3} \sin \vartheta$ and $\dfrac{Q l_x}{h^*} = \sin \vartheta$, and so expression (B.6) acquires the form:

$$\frac{S}{\hbar} = \frac{P_x}{2} \sin^2 \vartheta \qquad (B.7b)$$

Where the constant

$$P_x = \frac{\pi^3 \varepsilon_0 \varepsilon_{33}^f \hbar v_F}{e^2 \ln 3} \approx \frac{\pi^3 \varepsilon_0 \varepsilon_{33}^f \hbar v_F}{e^2} \quad (B.7c)$$

Hence the probability

$$w(\vartheta) \cong \exp(-P_x \sin^2 \vartheta) \quad (B.8)$$

Note, that the expressions (B.7)-(B.8) are actually $h^*$ independent.

Since the estimation for the exponential factor $P_x$ is typically essentially higher than unity (more than 100 for $\varepsilon \geq 500$), conduction per unit length of the p-n junction $\sigma_{pn}$ induced by a domain wall can be estimated as:

$$\sigma_{pn} = \frac{e^2}{\pi \hbar} \int_{-\vartheta_o}^{\vartheta_o} \frac{k_F w(\vartheta)}{2\pi} d\vartheta = \frac{e^2 k_F}{2\pi^2 \hbar} \int_{-\vartheta_o}^{\vartheta_o} \exp(-P_x \sin^2 \vartheta) d\vartheta \approx \Big|_{\sin^2 \vartheta \to \vartheta^2}$$

$$\approx \frac{e^2 k_F}{2\pi^2 \hbar} \int_{-\vartheta_o}^{\vartheta_o} \exp(-P_x \vartheta^2) d\vartheta = \frac{e^2 k_F}{2\pi^2 \hbar} \sqrt{\frac{\pi}{P_x}} Erf(\sqrt{P_x} \vartheta_o) \quad (B.9a)$$

Using that the wave vector $k_F \equiv \sqrt{\pi n_{2D}} \cong \sqrt{2\pi\gamma \frac{\varepsilon_{33}^f (P_S/e)}{\varepsilon_{33}^f + \gamma \varepsilon_g}}$ is controlled by the spontaneous polarization and permittivity only. Eq.(B.9a) can be rewritten as:

$$\sigma_{pn} = \frac{e^2 \sqrt{n_{2D}}}{2\pi \hbar} \frac{Erf(\sqrt{P_x} \vartheta_o)}{\sqrt{P_x}} \equiv \frac{e^2}{2\pi \hbar} \sqrt{2\gamma \frac{\varepsilon_{33}^f (P_S/e)}{\varepsilon_{33}^f + \gamma \varepsilon_d}} \frac{Erf(\sqrt{P_x} \vartheta_o)}{\sqrt{P_x}}, \quad (B.9b)$$

Since $P_x \approx \pi^3 \varepsilon_0 \varepsilon_{33}^f \hbar v_F / e^2$ in accordance with Eq.(B.7c) and the angle $\vartheta_0 \approx 1/\sqrt{P_x}$ is small enough, for estimates Equation (B.9b) can be simplified to the form:

$$\sigma_{pn} \cong \frac{e^2}{2\pi \hbar} \sqrt{\frac{n_{2D}}{P_x}} \approx \frac{e^2 \sqrt{n_{2D}}}{2\pi \hbar} \sqrt{\frac{e^2}{\pi^2 \varepsilon_0 \varepsilon_{33}^f \hbar v_F}} \equiv \frac{e^2}{\pi \hbar} \sqrt{\frac{\alpha}{\varepsilon_{33}^f} \frac{c}{\pi v_F} n_{2D}} \quad (B.10)$$

In Equation (B.10) we introduced the fine-structure constant $\alpha = e^2/4\pi\varepsilon_0(\hbar c)$, $c$ is the light velocity in vacuum.